\documentclass[sigconf]{acmart}
\setcopyright{none}
\renewcommand\footnotetextcopyrightpermission[1]{}
\usepackage{booktabs} 
\usepackage{graphicx}
\usepackage{xcolor}
\usepackage{hyperref}
\usepackage{booktabs}
\usepackage{graphicx}
\usepackage{multirow}
\usepackage{enumitem}
\usepackage[absolute,overlay]{textpos}

\copyrightyear{2026}
\acmYear{2026}
\setcopyright{cc}
\setcctype{by}
\acmConference[ICSE-NIER '26]{2026 IEEE/ACM 48th International Conference on Software Engineering}{April 12--18, 2026}{Rio de Janeiro, Brazil}
\acmBooktitle{2026 IEEE/ACM 48th International Conference on Software Engineering (ICSE-NIER '26), April 12--18, 2026, Rio de Janeiro, Brazil}
\acmPrice{}
\acmDOI{10.1145/3786582.3786846}
\acmISBN{979-8-4007-2425-1/2026/04}

\begin{document}

\title[SEALing the Gap: A Reference Framework for LLM Inference Carbon Estimation]{SEALing the Gap: A Reference Framework for LLM Inference Carbon Estimation via Multi-Benchmark Driven Embodiment}

\author{Priyavanshi Pathania$^\dagger$, Rohit Mehra$^\dagger$, Vibhu Saujanya Sharma$^\dagger$, Vikrant Kaulgud$^\dagger$, Tiffani Nevels*, Sanjay Podder$^\ddagger$, Adam P. Burden*} 
\affiliation{ 
	\institution{$^\dagger$Accenture Labs, India
		$^\ddagger$Accenture, India
		*Accenture, USA}
	\country{}
}
\email{{{priyavanshi.pathania, rohit.a.mehra, vibhu.sharma, vikrant.kaulgud}@accenture.com}}

\email{{{tiffani.y.nevels, sanjay.podder, adam.p.burden}@accenture.com}}
\renewcommand{\shortauthors}{Pathania et al.}

\begin{abstract}

Large Language Models (LLMs) are rapidly gaining traction in software engineering, yet their growing carbon footprint raises pressing sustainability concerns. While training emissions are substantial, inference quickly surpasses them due to the sheer volume of prompts processed. This shift underscores the urgent need for accurate, prompt-level carbon measurement during inference to enable informed, sustainability-focused decision-making. To address the limitations of existing approaches, in this paper, we outline the guiding principles for a novel \textit{reference} framework for LLM inference carbon estimation that can guide the design of future tools and provide a systematic foundation for advancing sustainability research in this domain. We also introduce \textit{SEAL}, an early embodiment of these principles that leverages a multi-benchmark-driven approach for per-prompt carbon estimation. Its initial validation shows promising results, positioning \textit{SEAL} as a foundation for standardized sustainability assessment across the LLM ecosystem.
 
\end{abstract}


\maketitle

\begin{textblock*}{6.5cm}(1.9cm,26.2cm) 
	Accepted for publication at ICSE 2026 (NIER Track)
	DOI: \url{https://doi.org/10.1145/3786582.3786846}
\end{textblock*}

\section{Introduction}\label{introductionv1}

With the proliferation of technologies such as coding assistants, generative AI, and agentic AI, traditional software engineering is undergoing a fundamental transformation. Recent industry surveys project that by 2026, 80\% of software vendors will have integrated generative AI capabilities into their software systems - an increase from less than 1\% in 2023 \cite{gartner2024cfo}. At the core of these technologies are LLMs, which are driving these unprecedented advances. On average, some LLMs process over 2.5 billion requests per day, fundamentally reshaping human-AI interaction, with natural language prompts emerging as the primary interface for interacting with AI systems \cite{NBERw34255}. However, while LLMs offer powerful capabilities, they have a significant carbon footprint of their own. This carbon footprint arises from the energy-intensive computational resources required to run them. For example, processing a single short prompt using OpenAI's GPT-4o consumes about 0.42 Wh of energy, which is equivalent to running a 10W light bulb for about 2.5 minutes; and this figure varies significantly with prompt length, semantic complexity, and response size \cite{jegham2025, husom2024price, poddar2025brevity}. As LLM adoption accelerates across software systems, the resulting growth in carbon emissions is projected to increase AI's share of global data center energy consumption from its current 5–15\% to as high as 25–50\% by 2030 \cite{KamiyaCoroama2025}. If left unaddressed, these rapidly growing emissions can have a serious impact on the sustainability of our environment.

This challenge underscores the urgent need for accurate and fine-grained (prompt-level) carbon measurement/estimation. In its absence, relevant stakeholders including developers, organizations, and policymakers, cannot effectively reason about the optimizations, interventions, trade-offs, or long-term sustainability strategies required to mitigate the carbon impact of LLM usage in software systems. 
Without reliable estimation, such trade-offs remain invisible, and decisions risk being guided solely by other metrics such as performance, cost, or latency, while silently accumulating environmental debt. Moreover, while carbon estimation is relevant across the entire lifecycle of LLMs, this paper focuses specifically on the inference phase, as the carbon cost of training and fine-tuning is quickly eclipsed by emissions generated during inference due to its widespread and repeated usage \cite{10.1145/3757892.3757901}.

While several approaches and tools exist for estimating energy consumption and carbon emissions, most target broader AI/ML systems rather than focusing specifically on LLM inference, which remains relatively new and inherently challenging due to the complexity of concepts such as tokens (input and output) and inference phases\footnote{Prefill refers to the phase where the LLM processes the input tokens to establish context, while Decode refers to the phase where output tokens are generated sequentially.} (prefill and decode). Recently, researchers and practitioners have begun to address LLM inference, however, existing efforts remain limited in scope and face significant challenges. These approaches are often highly approximate, coarse-grained, intrusive in nature, restricted to limited aspects of the LLM inference lifecycle (for example, decode-only), reliant on sub-standard/non-comprehensive ground truths, or neglect key LLM-specific attributes such as quantization or optimization, among other limitations. For example, CodeCarbon requires real-time access to the deployment environment of an LLM to gather the telemetry needed for estimating carbon emissions, an approach that is infeasible for closed-source LLMs provided by vendors such as OpenAI or Google \cite{codecarbon}. Taken together, these limitations highlight the absence of a reliable foundation for LLM inference carbon estimation. We believe that incremental improvements to existing tools are insufficient, as they remain fragmented, inconsistent, and often non-comparable across studies and deployments. What is needed is a \textit{reference} framework - a ``blueprint'' that establishes common guiding principles, against which future tools and methodologies can be developed, evaluated, and compared. It should address gaps in current approaches by explicitly articulating tool design expectations through a concise set of guiding principles tailored to LLM inference carbon estimation. We posit that such a \textit{reference} LLM inference carbon estimation approach should embody the following guiding principles (i) be non-intrusive (ii) provide fine-grained estimates (iii) be rooted in standard ground-truths (iv) leverage readily accessible attributes (v) support both open-source and closed-source models (vi) cover the entire inference lifecycle rather than only a specific phase, while demonstrating (vii) high accuracy. These are described in detail in the next section.

We believe that a fundamentally novel, multi-benchmark-driven approach to LLM inference carbon estimation can address these challenges and serve as the first concrete embodiment of the guiding principles of the \textit{reference} framework. While similar methodologies exist in related domains, such as ESAVE for estimating server and virtual machine energy consumption, no comparable approach currently exists for LLM inference carbon estimation \cite{pathania2022esave}.

In this paper, we motivate the need for a \textit{reference} carbon estimation framework for LLM inference. Next, we outline the guiding principles that any future tool in this space should follow to systematically address the limitations of existing approaches and provide a foundation for more reliable carbon estimation. Thereafter, we introduce, \textit{SEAL\footnote{SEAL=Sustainability Evaluations across the AI Lifecycle} for inference}, a novel multi-benchmark-driven approach, that embodies the \textit{"reference"} framework (follows all guiding principles) for estimating carbon emissions of LLM inference for both open and closed-source models, while supporting both the decode and prefill phases. \textit{SEAL for inference} is a non-intrusive approach that leverages a user-friendly set of configuration attributes and basic runtime telemetry to estimate the carbon emissions using custom-trained machine learning models, which have been trained on multiple industry-standard benchmarks. Early validations results appear promising. For brevity, we refer to \textit{SEAL for inference} simply as \textit{SEAL} throughout the remainder of this paper.

Thus, by reframing carbon estimation as a benchmarking problem, \textit{SEAL} serves as an early embodiment of our proposed \textit{reference} framework, highlighting how such guiding principles can open new directions for sustainability research in LLMs.
\section{Related Work}\label{relatedwork}

Carbon estimation for LLM inference has attracted increasing attention in recent years, and existing approaches can be broadly grouped into several categories. Monitoring approaches employ hardware or software profilers (such as nvidia-smi) or external power meters to capture inference energy consumption \cite{smi, samsi2023words, everman2023evaluating, henderson2020towards}. However, being intrusive, they are impractical for most production settings.

Simulation-based and workload-level approaches combine workload models and simulators with power models (such as GPU power curves) to estimate emissions for hypothetical/simulated workloads \cite{chien2023reducing}. However, these approaches rely heavily on workload assumptions, which often fail to generalize to real-world usage. Moreover, analytical approaches rely on predefined formulas based on proxies such as floating-point operations (FLOPs) or peak hardware efficiency, assuming a fixed linear relationship with energy usage \cite{faiz2023llmcarbon, desislavov2021compute, patterson2022carbon, towardsdatascienceblog, LaJavaness}. Such simplifications ignore factors such as I/O bottlenecks, utilization variability, and system overheads, leading to inaccurate estimates. Finally, dataset-driven approaches have recently emerged, which train ML predictors on benchmark-derived attributes to estimate inference energy \cite{ecologits}. Existing efforts remain limited in scope - typically relying on a single benchmark, using basic features, employing overly simplified models, and restricted to limited phases of the LLM inference lifecycle, thereby limiting their accuracy, generalizability, and adoption.

Taken together, these differences in assumptions, benchmarks, and evaluation scopes across existing approaches suggest that current efforts remain fragmented and ad hoc, rather than guided by a common foundation. This makes it difficult to design future carbon estimation tools that generalize beyond narrow settings. To move past this fragmentation, a principled \textit{reference} framework is needed - one that defines common guiding principles and offers a systematic basis for future LLM inference carbon estimation tools. Establishing such a foundation is the central focus of this paper.
\section{A "\textit{Reference}" Framework For LLM Inference Carbon Estimation} \label{motivation}

The limitations of existing approaches motivate the need for a \textit{reference} framework for carbon estimation of LLM inference. Such a framework must address the unique challenges posed by LLMs while overcoming the shortcomings of current tools and methodologies. Based on our research, we posit that such a \textit{reference} framework (refer Figure \ref{fig:referencearchitecture}) should satisfy the following guiding principles:

\begin{itemize}

	\item \textit{PA1: Non-intrusive}: The approach should not require invasive access to the deployment environment or low-level hardware telemetry. Current tools such as CodeCarbon demand real-time access to GPU utilization and power sensors, which is infeasible for closed-source LLMs and most production scenarios \cite{codecarbon}. A non-intrusive design ensures that the estimation remains practical across both research and production settings.

	\item \textit{PA2: Leverage Readily Accessible Attributes}: The framework should rely on readily available or easily measurable attributes such as model size, input token length, latency, or quality metrics, rather than restricted or opaque telemetry such as kernel settings, model architecture, or billing reports. For example, tools like Cloud Carbon Footprint needs access to the cloud service provider’s billing report, to estimate the carbon emissions based upon utilized cloud resources \cite{ccf}. Granting access to such confidential information might not be possible for an organization due to adherence to multiple security, privacy, and compliance-related protocols. Relying on accessible attributes lowers barriers to adoption and makes the approach more practical.
	
	\begin{figure}[t]
		\centering
		\includegraphics[width=1.0\linewidth]{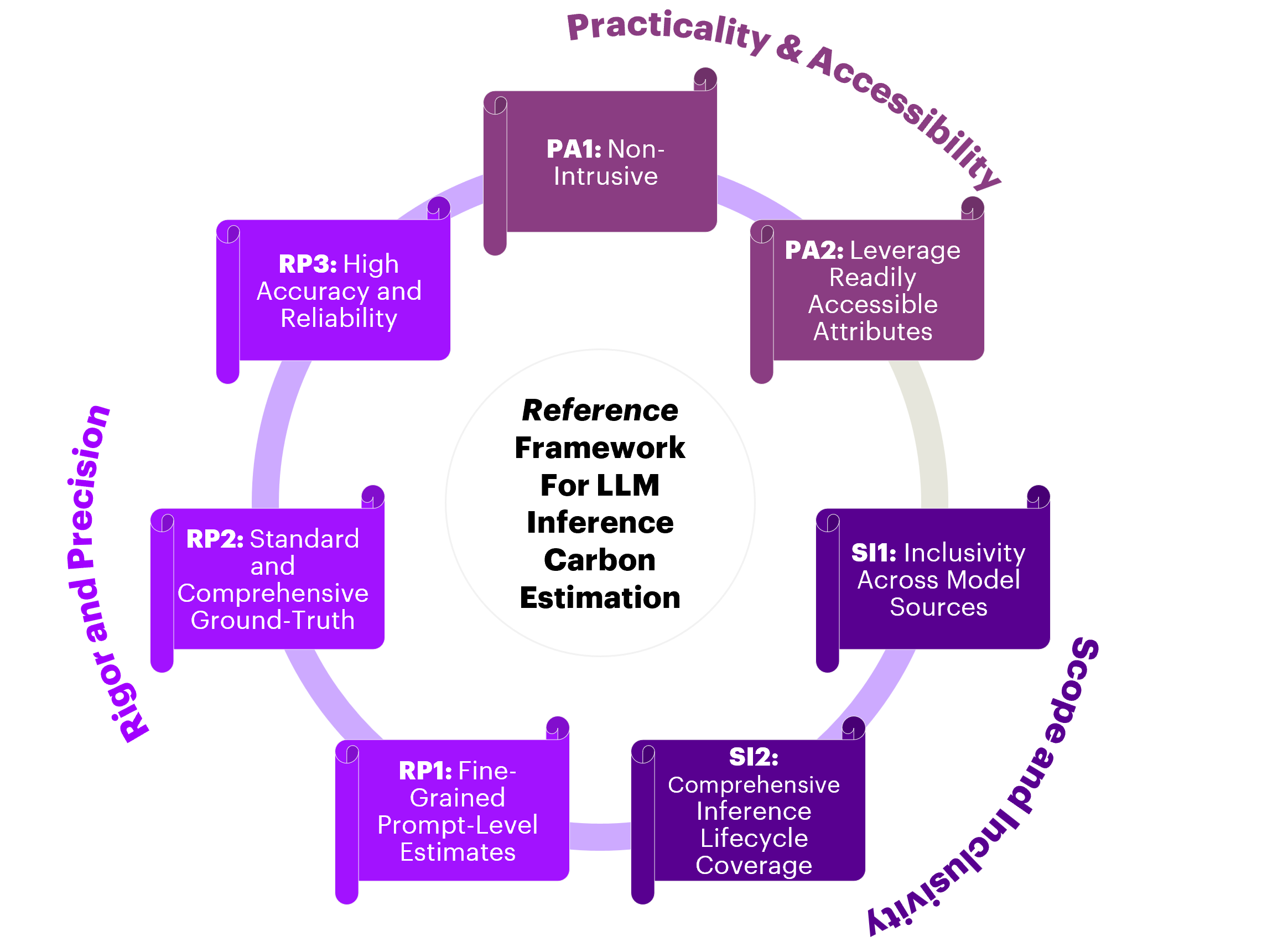}
		\caption{\textit{Reference} framework for LLM inference carbon estimation}
		\label{fig:referencearchitecture}
	\end{figure}

	\item \textit{SI1: Inclusivity Across Model Source}: A reference approach must work across both open-source LLMs (such as, Llama, Mistral, Qwen) and closed-source LLMs (such as, GPT-4o, Gemini 2.5 Pro, Claude). Many organizations rely on closed-source providers where low-level technical details are unavailable, so excluding them would make the approach incomplete and less practical for real-world adoption. For example, some recent simulation-based approaches estimate emissions by integrating GPU power models with workload simulators \cite{ozcan2025quantify}. While effective in controlled settings, these methods require access to detailed hardware performance curves and utilization patterns, which are not available for closed-source LLMs, limiting their practicality and adoption.

	\item \textit{SI2: Comprehensive Inference Lifecycle Coverage}: A reference approach should not be restricted to a single phase of inference, such as decode. Both the prefill and decode phases contribute significantly to energy consumption, especially for models with long context windows and applications like RAG, where the input context is often significantly larger than the number of output tokens generated \cite{wilkins2024offline, fernandez2025energy}. For example, tools like Ecologits that focus only on the decode phase would estimate the same carbon emissions regardless of whether the number of input tokens is 100 or 100 billion, which is highly inaccurate \cite{ecologits}. Ignoring any phase risks underestimating emissions and provides an incomplete picture for decision-making.
	
	\item \textit{RP1: Fine-Grained Prompt-Level Estimates}: Rather than relying on coarse-grained estimates, such as those at the level of entire data centers or virtual machines, the approach should provide estimates at a much finer granularity, ideally at the prompt-level. Fine-grained estimates enable developers and organizations to quantify the benefits of different trade-off strategies. For example, in a carbon-efficient retrieval-augmented generation (RAG) pipeline, the decision of how much context to include in a prompt becomes a quantifiable trade-off, as per-prompt estimates directly reveal the carbon impact of alternative strategies.
	
	\item \textit{RP2: Standard and Comprehensive Ground-Truth}: Estimates should be rooted in widely recognized, reproducible, and comprehensive ground-truth data sources, such as public and open-source industry-standard benchmarks. Without such standardization, estimates remain non-comparable across studies, tools, or organizations, undermining their credibility and practical adoption. For example, some prior works derive carbon estimates using proxies like floating-point operations (FLOPS) or theoretical peak hardware efficiency, assuming a fixed linear mapping to energy usage \cite{faiz2023llmcarbon}. Such simplifications ignore real-world factors like I/O bottlenecks, utilization variability, or overheads, leading to inaccurate estimates. In contrast, standardized benchmarks reflect representative operating conditions and offer a more reliable picture when comparing different models or deployment settings.
	
	\item \textit{RP3: High Accuracy and Reliability}: Finally, the framework must demonstrate consistently low error margins when validated against external datasets or independent measurements. High accuracy is essential for building trust among relevant stakeholders, who may base optimization and governance decisions on these estimates.
	
\end{itemize}

Collectively, these guiding principles articulate the foundations of a \textit{reference} framework for LLM inference carbon estimation, addressing the shortcomings of current methodologies while providing a systematic blueprint for future tools and approaches. The broader aim of this paper is to initiate community discussion around these principles and establish a common foundation to guide the design of future estimation tools in this space.
\section{An Embodiment of our \textit{Reference} Framework - \textit{SEAL For Inference}}\label{sealv1}

To demonstrate how the guiding principles of our \textit{reference} framework can be embodied in practice, we developed \textit{SEAL for inference}. It is an early implementation that highlights how a multi-benchmark-driven approach can address the limitations of existing approaches and represents a promising step toward a standardized approach for prompt-level carbon estimation during inference. Please note that the implementation focuses on estimating the energy consumption, which can then be converted to carbon emission estimates using region-specific carbon intensities \cite {ccf}. \textit{SEAL} contributes as an early embodiment of the proposed reference framework and demonstrates, in a novel manner, how prompt-level inference energy estimation can be realized in alignment with the articulated guiding principles.

For building a non-intrusive approach, we focused on estimation rather than monitoring to avoid intrusive access to the deployment environment (\textbf{PA1}). To realize this, we conducted a landscape study of existing LLM benchmarks to identify suitable candidates to serve as standardized ground truths (\textbf{RP2}). Benchmark(s) selection was guided by four key questions (i) are there benchmarks that capture energy/carbon at the prompt level? (\textbf{RP1}) (ii) has the benchmarking process been conducted on standard hardware configurations to enable extrapolation to real-world scenarios? (iii) is the benchmark sufficiently large to support training a generalized ML model? and (iv) can the attributes/keys in the benchmark (such as LLM name) be linked with other standard benchmarks for building a comprehensive and feature-rich dataset (such as those concerning model quality or performance)? Importantly, such linkage is only feasible when methodologies either align or concern orthogonal dimensions: for example, two energy benchmarks cannot be combined if their energy measurement methods diverge, whereas energy and quality benchmarks can be jointly analyzed since they address distinct concerns. Multiple benchmarks were preferred over a single one to enhance feature richness and enable analysis of how orthogonal features influence energy. This novel strategy was inspired by the work of Vasilescu et al., who merged GitHub and Stack Overflow datasets to study cross-platform developer activity \cite{6693332}.

Several benchmarks fulfilled these criteria, but for our initial explorations, we selected two open-source benchmarks: the LLM-Perf Leaderboard (3173 observations) and Open LLM Leaderboard (2045 observations) \cite{llm-perf-leaderboard, open-llm-leaderboard-v2}. The LLM-Perf Leaderboard reports the per-prompt, per-token, phase-wise performance (such as latency, throughput, and memory consumption) and energy consumption of LLMs across diverse hardware and software setups, while the Open LLM Leaderboard evaluates an overlapping set of LLMs on six widely-used quality benchmarks, such as MMLU-Pro and BBH \cite{wang2024mmlu, suzgun2022c}. By merging these benchmarks (using a combination of model name and precision as unique key), we were able to construct a comprehensive dataset (3042 observations) that captured both energy and quality characteristics of models, for training \textit{SEAL}.

After initial data cleanup, feature engineering, and extensive exploratory data analysis, we selected seven features to train \textit{SEAL}. These features were chosen based on their (i) contextual availability for a given prompt (number of input and output tokens), (ii) public availability (model size), (iii) measurability via public or private LLM APIs (latency per input and output token), (iv) assumability (underlying GPU), and (v) correlation with energy (quality features such as BBH and MMLU-Pro) (\textbf{PA2}) (\textbf{SI1}). While we acknowledge that features like model size, GPU, and latency per token are not always directly available for vendor-deployed models, they can often be approximated reasonably. For example, model size and GPU from reputable academic/public sources, latency via vendor APIs, and GPU through popularity-based assumptions. Although not perfectly accurate, these approximations still provide researchers and practitioners with valuable insights to support sustainability decisions that might otherwise be impossible. Finally, while other features such as quantization, attention, and kernel configurations also influence energy, they were excluded due to their limited availability and because their effects are indirectly reflected in latency, which we already included as a training feature.

Since our selected features were primarily continuous and categorical, we employed a set of 13 regression-based algorithms to train our estimation models, and evaluated their goodness of fit using the mean absolute percentage error metric (MAPE). Notably, we trained two separate models for both the Prefill and Decode phases (\textbf{SI2}). Because our dataset contained model sizes ranging between 0-111B parameters, we developed one model for interpolation (estimating within this range) and another for extrapolation (estimating beyond this range). This distinction was necessary because many currently available LLMs are larger than 111B parameters, making extrapolation particularly important, and regression models optimized for interpolation generally perform poorly on extrapolation tasks (e.g., gradient-boosted decision trees \cite{8999249}).

\begin{table}[t]
	\ttfamily
	\centering
	\caption{10-fold cross-validation results of \textit{SEAL’s} early prompt-level energy estimation models across interpolation and extrapolation scenarios.}
	\label{tab:my-table}
	\resizebox{\columnwidth}{!}{%
		\begin{tabular}{@{}lllll@{}}
			\toprule
			\textbf{Phase}           & \textbf{Generalization Type} & \textbf{MAPE \%}    & \textbf{$R^2$} & \textbf{Regressor Type} \\ \midrule
			\multirow{2}{*}{Decode}  & Interpolation                & $6.98\% \pm\; 0.56\%$  & 0.999       & XGBoost Regressor       \\
			& Extrapolation                & $31.58\% \pm\; 2.78\%$ & 0.986       & Ridge Regressor         \\ \midrule
			\multirow{2}{*}{Prefill} & Interpolation                & $5.36\% \pm\; 0.46\%$  & 0.995       & XGBoost Regressor       \\
			& Extrapolation                & $24.85\% \pm\; 1.77\%$ & 0.994       & Ridge Regressor         \\ \bottomrule
		\end{tabular}%
	}
\end{table}

\begin{table}[]
	\ttfamily
	\centering
	\caption{Early results of our external validation experiments using published empirical data \cite{wilkins2024offline}}
	\label{tab:my-table2}
	\resizebox{\columnwidth}{!}{%
		\begin{tabular}{@{}lllll@{}}
			\toprule
			\textbf{LLM} & \textbf{Active Parameters} & \textbf{Energy From Empirical Data} & \textbf{Energy Estimated Using \textit{SEAL}} & \textbf{Error \%} \\ \midrule
			llama-2-7B   & 7 B                        & 349.96 Joules                         & 425.60 Joules                          & 19.51\%       \\
			llama-2-13B  & 13 B                       & 602.27 Joules                         & 707.20 Joules                         & 16.02\%       \\ \bottomrule
		\end{tabular}%
	}
\end{table}

Out of the 13 regression algorithms we tested, XGBoostRegressor (an optimized gradient boosting algorithm) achieved the lowest MAPE in interpolation scenarios, whereas Ridge Regressor (a linear model with regularization) achieved the lowest MAPE in extrapolation scenarios. Table 1 summarizes the 10-fold cross-validation results for both scenarios. We also performed early external validation against empirical data from Wilkins et al. for two models (LLaMA-2-7B and LLaMA-2-13B) using a prompt with 38 input tokens and 64 output tokens \cite{touvron2023llama, wilkins2024offline}. Table 2 presents the results of this external validation, highlighting an average error of 17.76\% (\textbf{RP3}). Overall, both cross-validation and early external validation results appear promising. Further comprehensive evaluations in this direction represent a major portion of our planned future work.

In essence, \textit{SEAL} demonstrates how a multi-benchmark-driven approach can embody all the guiding principles of our \textit{reference} framework: it is non-intrusive, requiring only attributes such as tokens, latency, and model size rather than restricted telemetry; it provides fine-grained prompt-level estimates by modeling energy at the per-prompt level; it is grounded in standardized benchmarks by merging LLM-Perf and Open LLM Leaderboard; it leverages readily accessible attributes that can be measured or reasonably approximated even for vendor models; it ensures inclusivity by supporting both open- and closed-source models; it captures the inference lifecycle through separate models for prefill and decode; and it demonstrates early evidence of accuracy and reliability through promising cross-validation and external validation results. While still early, \textit{SEAL} offers a concrete foundation that the community can build upon to advance toward standardized practices in LLM inference carbon estimation. The proposed impact of this work lies not in immediate state-of-the-art replacement, but in providing a common reference point that can influence how future LLM inference carbon estimation approaches are designed and evaluated.

\section{Future Plans}\label{futurework}

Looking ahead, we intend to advance the \textit{reference} framework by evaluating its completeness, identifying missing guiding principles, and refining it into a robust foundation. A key direction will be to articulate what a standardized reference ``benchmark'' should look like and how it should be built to support the development of more accurate and widely adoptable estimation tools.

On the \textit{SEAL} front, we plan on exploring advanced modeling techniques to further enhance estimation accuracy, incorporating additional benchmarks to capture a broader range of energy-influencing features, and undertaking comprehensive external validations. In particular, collaborations with LLM vendors and cloud providers could enable systematic evaluation of closed-source deployments, strengthening the generalizability and trustworthiness of the approach. Furthermore, integrating \textit{SEAL} into developer-facing tools can enable real-time, sustainability-aware decision support, with the long-term goal of evolving this early-stage research into a community asset for both academic and industrial adoption.

\bibliographystyle{ACM-Reference-Format}
\bibliography{acmart}

\end{document}